\documentclass[sigconf]{aamas} 
\renewcommand\footnotetextcopyrightpermission[1]{} 
\usepackage{hyperref}
\usepackage{bbm}
\usepackage{amsfonts}
\usepackage{subcaption}
\usepackage[linesnumbered,ruled,vlined]{algorithm2e}
\usepackage{balance} 

\usepackage{enumerate} 
\SetKwInput{KwInput}{Input}                
\SetKwInput{KwOutput}{Output}              

\copyrightyear{2022}
\acmYear{2022}
\acmDOI{}
\acmPrice{}
\acmISBN{}

\acmSubmissionID{383}

\title[AAMAS-2022]{Multi-Agent Adversarial Attacks for Multi-Channel Communications}

\author{Juncheng Dong}
\affiliation{
  \institution{Duke University}
  \city{Durham}
  \country{U.S.A}}
\email{juncheng.dong@duke.edu}

\author{Suya Wu}
\affiliation{
  \institution{Duke University}
  \city{Durham}
  \country{U.S.A}}
\email{suya.wu@duke.edu}

\author{Mohammadreza Soltani}
\affiliation{
  \institution{Duke University}
  \city{Durham}
  \country{U.S.A}}
\email{mohammadreza.soltani@duke.edu}

\author{Vahid Tarokh}
\affiliation{
  \institution{Duke University}
  \city{Durham}
  \country{U.S.A}}
\email{vahid.tarokh@duke.edu}

\begin{abstract}
Recently Reinforcement Learning (RL) has been applied as an anti-adversarial remedy in wireless communication networks. However studying the RL-based approaches from the adversary’s perspective has received little attention. Additionally, RL-based approaches in an anti-adversary or adversarial paradigm mostly consider single-channel communication (either channel selection or single channel power control), while multi-channel communication is more common in practice. In this paper, we propose a multi-agent adversary system (MAAS) for modeling and analyzing adversaries in a wireless communication scenario by careful design of the reward function under realistic communication scenarios. In particular, by modeling the adversaries as learning agents, we show that the proposed MAAS is able to successfully choose the transmitted channel(s) and their respective allocated power(s) without any prior knowledge of the sender strategy. Compared to the single-agent adversary (SAA), multi-agents in MAAS can achieve significant reduction in signal-to-noise ratio (SINR) under the same power constraints and partial observability, while providing improved stability and a more efficient learning process. Moreover, through empirical studies we show that the results in simulation are close to the ones in communication in reality, a conclusion that is pivotal to the validity of performance of agents evaluated in simulations.
 
\end{abstract}

\keywords{RL, Communication, Multiagent System, Adversary Attacks}

\newcommand{\BibTeX}{\rm B\kern-.05em{\sc i\kern-.025em b}\kern-.08em\TeX}

\begin{document}


\pagestyle{fancy}
\fancyhead{}

\maketitle 


\section{Introduction}
In recent years, we have witnessed the success of reinforcement learning (RL) in many sequential decision-making problems in machine learning, optimal control, and wireless communication~\cite{Mao2018,Ye2020,li2018deep}. This is mainly due to the use of deep neural networks (DNNs) and their expressive power in the traditional RL algorithms such as Deep Q-Network (DQN), Double DQN (DDQN), and Deep Deterministic Policy Gradient (DDPG)~\cite{DQN,DDQN,DDPG}. However, many practical sequential decision-making problems involve modeling the interaction between multiple agents and the environment. For instance, defending wireless communication networks against adversary attacks typically requires the collaboration between different nodes in the network. Wireless networks are highly vulnerable to adversarial attacks due to the open and sharing nature of the wireless medium. One of the challenges in these networks is to detect/predict adversarial activities and designing proper defense mechanisms, which typically require cooperation/competition among multiple friendly or adversary agents~\cite{Aref2017,Zhou2021}. Early attempts to prevent the adversarial attacks in wireless communication included the frequency hopping techniques. By choosing transmission frequency from a set of available frequency chunks (either randomly or by using some learning procedure) the defender attempts to avoid the compromised frequencies~\cite{prime_antijamming, aj_survey}.
\begin{figure}[h]
    \centering
    \includegraphics[width=\linewidth]{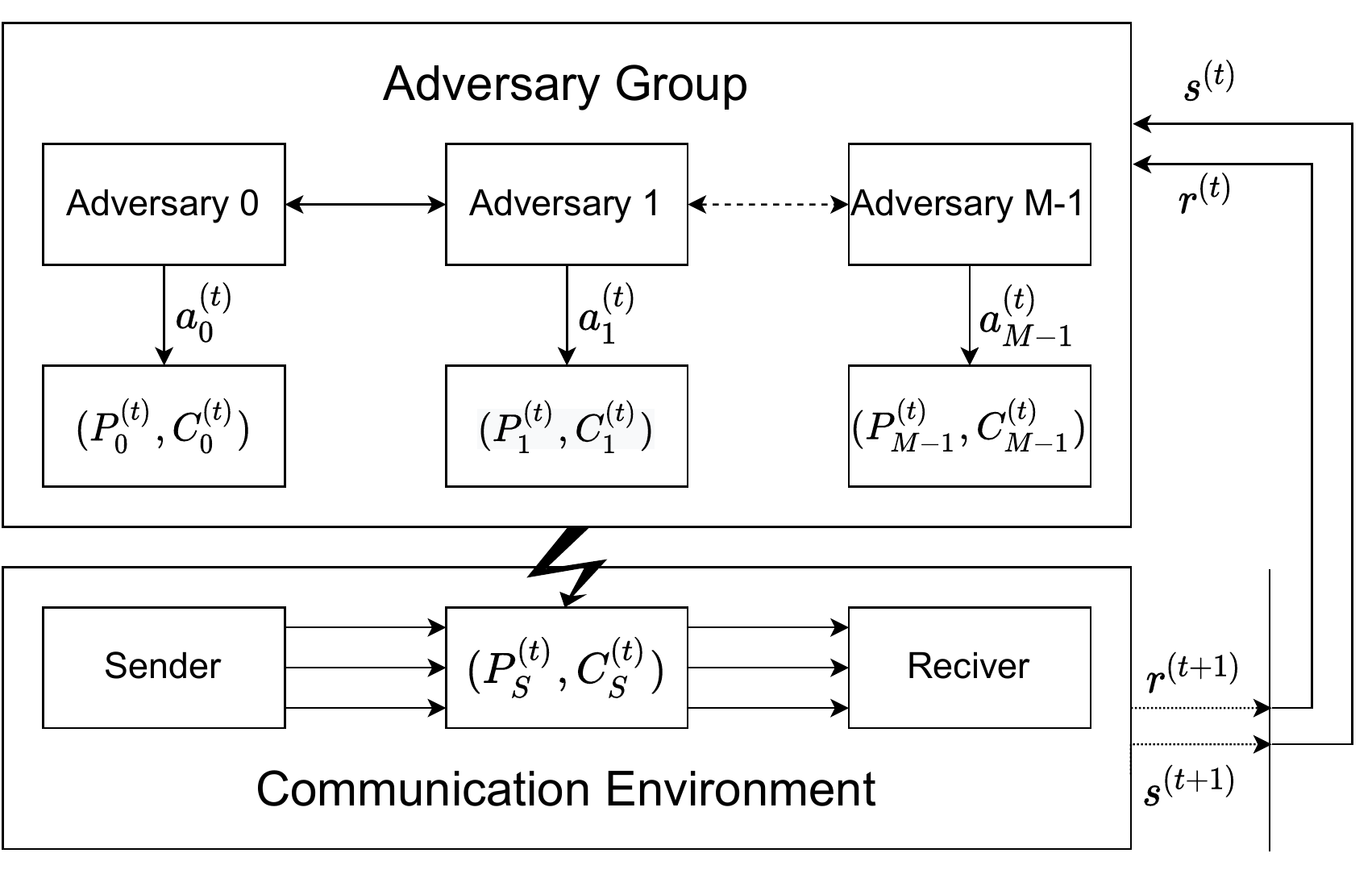}
    \caption{Multi-channel wireless communication system with $M$ adversaries and one pair of sender/receiver. At time $t$, both sender/receiver and adversaries select a channel and a power level respectively as $\left(P_S^{(t)}, C_S^{(t)}\right)$ and $\left(P_j^{(t)}, C_j^{(t)}\right)$. Adversaries observe state $s^{(t)}$ and receive rewards $r_j^{(t)}$ from communication environment.}
    \label{fig:system_model}
\end{figure}
Recently RL has been successfully applied for designing algorithms for defending against adversary attacks in a hostile wireless communication environment~\cite{Ye2020,Liu2018,Xu2020,Chen2018}. However, using the RL-based methods from the adversary's perspective has received little attention. This is especially important since a better understanding of adversaries' behavior can lead to better design of defense mechanisms as well as active defense. Furthermore, the current endeavor for utilizing the RL-based methods either in anti-adversary or adversary paradigm mostly consider communication within a single channel (either channel selection or single channel power control), while multi-channel communication is more common in real scenarios. 

In this paper, we propose a multi-agent adversary system (MAAS) based on RL for modeling and analyzing adversaries (e.g., jammers) in a wireless communication scenario by a careful design of the reward function for realistic multi-channel communication scenarios. Through extensive simulations, we show that the proposed MAAS learns to choose the transmit channel and the most efficient power allocation for attack without any prior knowledge of the sender strategy. In particular, our results demonstrate that using MAAS and implicit collaboration between the adversaries can provide better performance and success rates compared to the single-agent adversary (SAA) case. In addition, MAAS is fault-tolerant and robust to the failure of some agents since other agents may take over the failed agent's duty and continue their operations. Our contribution can be summarized as follows: 
\begin{itemize}
    \item We use an RL-based approach for modeling and analyzing the adversaries in a multi-channel and realistic wireless communication scenario.
    \item  We show the superiority of multi-agent adversaries versus the single-agent by achieving a significant reduction in the signal-to-interference-plus-noise ratio (SINR) under the same power constraint and partial observability. Specifically, we show that MAAS is able to learn more efficiently and stably by reducing the exponential complexity of SAA's action space to a linear complexity through distributing actions to multi-agents.
\end{itemize}

The rest of this paper is organized as follows. In Section 2, we define our system model of communication, provide some background about RL, and briefly discuss existing work on RL-based anti-adversaries. In Section 3, we propose and discuss the MAAS in detail together with studying the multi-channel system model and the advantages of using multi-agents instead of single-agent. In Section 4, we present simulation results on various types of sender strategy and demonstrate the superior performance of the proposed MAAS against baseline approaches, including the single-agent case, random adversaries, and greedy adversaries. We conclude in Section 5.

\section{Backgrounds}
In this section, we first review some required background for the single-agent RL as well as multi-agent RL, and then provide some related works of using RL techniques and multi-agent systems in the wireless communication. 

\subsection{Single-agent RL}
A single-agent RL is an agent that learns how to make sequential decisions by interacting with the environment through receiving a feedback signal called \emph{reward}. To choose the best action in each time step, the agent tries to choose an action, maximizing its long-term (discounted) rewards. In this scenario, the environment is modeled by a Markov Decision Process (MDP)~\cite{RLBible} defined by a 5-tuple of $(\mathcal{A}, \mathcal{S}, p, R, \gamma)$, where $\mathcal{A}$ and $\mathcal{S}$ denote the action space and state-space respectively. $p$ is the transition probability, $R$ denotes the reward function and $\gamma$ is the discount factor that trades off the immediate and future rewards. In a single-agent RL, at time step $t$, the agent chooses its action $a_t$ according to a \emph{policy} function $\pi$. As a result of choosing the action $a_t$, the environment changes its current state $s_t$ to the next state $s_{t+1}$, and returns a real-valued reward $r_t$ (feedback) to the agent. Hence, the interaction of an agent with an environment modeled by the above MDP is defined as follows:
\begin{enumerate}
    \item The state of the environment at time $t$: $s_t \in \mathcal{S}$
    \item Agent's action at time $t$: $a_t \in \mathcal{A}$
    \item State transition probability: $p: \mathcal{S}\times \mathcal{A} \rightarrow \mathcal{S} $
    \item Reward function: $R: \mathcal{S}\times \mathcal{A} \rightarrow \mathbb{R}$
    \item Agent's policy: $\pi: \mathcal{S} \rightarrow \mathcal{A} $
\end{enumerate}
Here, policy $\pi$ is a mapping from the current state $s_t$ to the current action $a_t$, i.e., $\pi=\pi(a_t|s_t)$. Moreover, $p$ captures the dynamics of the environment and is written as a conditional probability of next state given the current state and action, that is $p(s^{(t+1)}|s^{(t)}, a^{(t)})$. Also, the reward function $R$ gets the current state and action as input and returns the agent's reward, i.e. $r^{(t)} := R(s^{(t)}, a^{(t)})$. As mentioned above, the objective function of a single-agent RL is to maximize the discounted cumulative rewards by selecting the best policy $\pi$. That is,
\begin{align*}
    \max_{\pi}\mathbb{E}\Big{[}\sum_{t=0}^{\infty}\gamma^t R^{(t)} ;\pi\Big{]}.
\end{align*}
If the state transition probability and the reward function are known (i.e., planning problem), then the optimal policy $\pi^*$ can be solved through iterative algorithms such as the value iteration algorithm~\cite{RLBible}. However, in many applications, the dynamics of the environment, $p$, and the reward function $R$ are not known. Model-free algorithms~\cite{DQN} are RL algorithms that solve MDP without requiring to know $p$ or $R$ in advance. One of the most widely used model-free RL algorithm is the Q-learning algorithm that finds the optimal policy $\pi^*$ by maximizing action-value (also called Q-function) $Q^{\pi}(s_t, a_t) = \mathbb{E} \Big{[}\sum_{t=0}^{\infty}\gamma^t R^{(t)} \rvert s_t, a_t;\pi\Big{]}$. The action-value function returns the expected discounted cumulative reward when an agent takes an action $a_t$ in state $s_t$, and following the policy $\pi$ afterward. As a result, the optimal policy $\pi^*$ is the one that maximizes $Q(s_t,a_t)$ for all $s_t$ and $a_t$. That is, 
\begin{align*}
    \pi^* = \operatorname*{argmax}_{\pi} Q^{\pi}(s_t, a_t).
\end{align*}
The classical Q-learning algorithm maintains a Q-table to store the corresponding Q-values for all state-action pairs. At each time step $t$, values in Q-table is updated based on the Bellman optimally equation,  $Q^{\pi^*}(s_t,a_t) = r^{(t)} + \gamma\mathbb{E}[ Q^{\pi^*}(s_{t+1},a_{t+1})]$. Values in Q-table converges to the optimal value $Q^{*}(s,t)$ after enough iterations. However, for many problems, the state space grows exponentially with the dimension of state and the state space can often be continuous; hence, making the Q-learning impractical. To get around these issues, deep Q-learning applies a neural network to estimate the Q-values and to replace the Q-table.

\subsection{Multi-agent RL (MARL)}
Existing Multi-agent RL algorithms can be categorized by how the agents are trained and how they make a decision after training (execution). Both training and execution schemes can be divided into centralized and distributed approaches~\cite{MARL_survey}. Distributed adversaries are usually more robust than the centralized ones when there is a counterattack from the anti-adversary side, or when some adversaries fail to operate properly. On the other hand, distributed training/execution can be expensive and unstable in a hostile environment due to the communication overhead between agents. As a result, distributed training and execution are more desired for multi-agent adversary systems. We refer the readers to~\cite{oroojlooyjadid2021review, MARL_survey} for a detailed discussion about MARL.

\subsection{Applications of RL in Communication}
The progress in the field of RL spurred a new interest in intelligent wireless communication and networking~\cite{Mao2018, Luong2019}. The rapid proliferation of security concerns has made considerable attention for investigating adversary attacks and mitigating the jamming interference, resulting in extensive studies in intelligent anti-adversary (anti-jamming) techniques~\cite{Pirayesh2021}. Previous methods have used the game theory to find the optimal anti-jamming strategies~\cite{Wu2012, Hanawal, Gwon2013, Wang2011, Tang2016}. Most of these theoretical studies formulated the interaction of a sender with a jammer/adversary as a zero-sum Markov game~\cite{Wu2012, Hanawal, Gwon2013} or a Stackelberg game~\cite{Tang2016}. The two players in the game dynamically decide actions to optimize the opposite objectives. The actions typically include which channel to switch and what transmission rate to use. Then the optimal strategy of the sender can be derived by applying theoretical results from the MDP and the prior knowledge about jammers. In these methods, the transition probability is necessary for such a derivation; therefore, it cannot be used in more challenging and realistic scenarios.

Since RL methods are an appealing framework for selecting an optimal action that yields the highest reward for a given state, they have been used extensively in modeling many anti-adversary wireless communication systems, including cognitive radio networks~\cite{Han2017, Shi2018, Liu2018, Xu2020, Ye2020}, in-air unmanned aerial vehicular (UAV)~\cite{Lv2017, Xiao2018, Lu2020}, Internet of Things (IoT) services~\cite{Chen2018}. For example,~\cite{Han2017} has proposed a deep Q-learning-based anti-jamming strategy for cognitive radio networks. In this system, the secondary users (SUs), learned by convolutional neural networks (CNN) choose a frequency channel that avoids interference with the primary users (PUs). Along the same line, an anti-jamming deep reinforcement learning algorithm (ADRLA) has been proposed by~\cite{Liu2018} based on recursive convolutional neural networks (RCNN). For a more realistic scenario with a large state and action space, some authors incorporate the power allocation into users' decision-making by utilizing advanced DDQN~\cite{Bi2019, Xu2020} and fast convergence investigation~\cite{Ye2020}. To adapt to the multi-agent environment, a sequence of multi-agent reinforcement learning (MARL)-based schemes have been designed to find anti-jamming policies in multi-channel systems~\cite{Aref2017, Yao2019, Elleuch2021, Zhou2021}. In addition to the RL-based methods,~\cite{Erpek2019} has used a deep classifier-based jamming attack and anti-jamming defense to distinguish the success of attacks from the genuine transmissions by deep classifiers.

Regarding the jamming strategies, there are a few works about anti-jamming strategies in the presence of intelligent attacks~\cite{Arjoune2020}. Most of the existing anti-jamming schemes are developed to confront conventional types of jammers, for example, random jammers, sweeping jammers, and reactive jammers~\cite{Erpek2019}.

In this paper, we propose a MARL-based system for multi-channel communications from the adversary/jammer perspective, where the jammers are smart and are able to adapt to a multi-channel environment efficiently, especially when current communication networks tend to be decentralized and ad-hoc~\cite{Luong2019}. 
\section{Multi-agent Adversary System (MAAS)}
In this section, we first describe our proposed system model for simulating multi-channel communication. Based on the system model, reward functions are carefully designed for the correct guidance of MAAS. We also discuss in detail the learning process of MAAS. Figure~\ref{fig:MAAS} illustrates the  workflow of our proposed MAAS and algorithm~\hyperref[alg:MAAS]{1} presents the pseudo-codes for training and prediction of MAAS.
\subsection{System Model of Communication}
We now define our communication model. In our scenario, we assume there is a pair of sender/receiver and there are $N$ available channels for sending the signal (see figure \ref{fig:system_model}). We assume that the communication between the sender and the receiver and also the adversaries happens only at discrete time steps  $t=0,1,2,...$. At each time step $t$, the sender is allowed to choose multiple channels to send its signal to the receiver. The amount of power allocated to each channel is assumed to be fixed during the time that channel is used for communication. We also assume that there are $M$ adversaries (indexed by the index $j = 0,1,...,M-1$), each selects one channel out of $N$ available channels and a power level from a set $P=\{P_0,P_1,\ldots,P_K\}$, where $0 \leq P_0 < P_1<\dots<P_K$). This means that the adversary $j$'s action at time $t$ is a 2-dimensional vector $a_j^{(t)} = [C_j^{(t)},P_j^{(t)}]^T$, where $C_j^{(t)}$ is the channel selected by adversary $j$, and $P_j^{(t)} \in P$ is its selected power level at time $t$. 

In our scenario, the goal is that adversaries can select their actions to maximize decrease of the quality of communication (QoC) between the sender and the receiver. The (QoC) at each time step $t$ is defined by the signal-to-interference-plus-noise ratio (SINR):
\begin{align}
\label{sinr}
    \text{SINR}^{(t)} = \frac{P_{S}^{(t)}*h_S}{\eta+\sum_{j=0}^{M-1}P_{j}^{(t)}*h_j*I(C_{j}^{(t)}=C_{S}^{(t)})},
\end{align}
where $\eta$ denotes the communication noise, and $h_S$ and $h_j$ are power gains for the sender and adversary $j$, respectively. 

In order to evaluate the performance of an adversary in our communication model, we define \emph{Success of Attack} (SA) as a binary-valued function as follows:
\begin{equation}
    SA = \mathbbm{1}{ ( SINR^{(t)} < \tau SNR^{(t)})},
\end{equation}
where $\tau$ denotes a pre-defined threshold, and SNR (signal-to-noise-ratio) is the maximal achievable SINR for the receiver, calculated as $SNR = \frac{P_s * h_s }{\eta}$ (i.e., the SINR without any interference). Also, $\mathbbm{1}(condition)$ denotes the indicator function defined as 1 if the condition is true, and 0 otherwise. We also define the \emph{Success Rate of Attack} (SRA) as the ratio of the number of SA over a time interval $T>0$:
\begin{align}
    \text{SRA} = \frac{\sum_{t=0}^{T-1}\mathbbm{1}{ ( SINR^{(t)} < \tau SNR^{(t)} ) }}{T}.
\end{align}
The choice of $\tau$ is problem-specific. However, as we will see in the experimental results, the change of $\tau$ will not have a dramatic effect in changing the SRA value of the proposed MASS. 

\subsection{Design of Reward Function}
In RL, designing a reward function is a crucial step as it is the only feedback that an agent can receive from the environment; as a result, it has a huge effect on learning the best policy~\cite{RLreward} and improving the overall performance. A standard choice for the reward function in many anti-adversary RL-based approaches is given by the SINR. However, directly using the SINR as the reward function may be misleading since it may increase the reward function artificially; thus, achieving a higher SRA for the MAAS compared to the single-agent systems. To overcome this issue, our reward function includes two parts: the portion of channels blocked by adversaries, and the power cost incurred by adversaries for attacking the communication between the sender and the receiver.
\begin{itemize}
    \item\textbf{Channels blocked by adversaries.} This can be computed by the decrease in the Shannon channel capacity calculated as $B*\left(\log_{2}(1+\text{SNR}^{(t)}) - \log_{2} (1+\text{SINR}^{(t)})\right)$. 
    \item\textbf{Power cost.} The reward function should include a term, indicating the cost of power to penalize adversaries if they use the allocated power inefficiently. We consider a constant power cost, $Cost_{power}$ for all the adversaries and assume that the cost is known to adversaries throughout the communication.
\end{itemize}
Hence, the total reward function for the whole multi-agent system is given by:
\begin{align}\label{teamreward}
   &R^{(t)} = B*\left(\log_{2}(1+\text{SNR}^{(t)}) - \log_{2} (1+\text{SINR}^{(t)})\right)\nonumber\\ &\hspace{3.7cm}-C_{\text{power}}*\sum_{j=0}^{M-1}P_j^{(t)},
\end{align}
where $P_j$ is the power used by adversary $j$.
\subsection{RL for Multi-Agent Adversary System} 
Unlike the episodic RL problems in which there is a terminal state indicating the end of one episode (e.g., game playing~\cite{MADDPG}), modeling adversaries in wireless communication requires continuous learning. Due to the continuous state and non-Markovian dynamic of environment caused by simultaneous learning of multi-agents as well as the requirement for continuous learning, we choose the Double Deep Q-learning with prioritized experience replay as RL agent for each adversary in MAAS for faster adaption to the environment~\cite{DDQN, PriorizedER}. Next, we describe in detail the learning process.

\textbf{State.} Similar to existing works~\cite{Ye2020}, we use the SINR at time $t-1$ as the state for the environment at time $t$. That is, $s^{(t)}= SINR^{(t-1)}$ for $t=1,2,3,\ldots.$ We note that in wireless communication, the true value of SINR is not known for any agent other than the receiver. Here, we first follow the current practice for using the expected SINR for the rewards. Later, we show through simulations that the success rate of adversaries without using the true value of SINR stays at the same level as the former case. To our knowledge, conducting this set of simulations is novel and shows the stability as well as reliability of the performance of the proposed RL method.

\textbf{Reward.} In addition to the multi-agent system reward defined in equation~(\ref{teamreward}), a learning agent (e.g., adversary $j$ for $j=0,1,\dots,M-1$) in MARL can have its own reward for distributed training at time $t$ defined as follows:

\begin{align}\label{eq:reward}
    &R_j^{(t)} = B*\left(\log_{2}(1+\text{SNR}^{(t)}) - \log_{2}(1+\text{SINR}^{(t)})\right)\nonumber\\
    &\hspace{3.7cm}-Cost_{power}*P_j^{(t)}.
\end{align}

\begin{figure}[h]
    \centering
    \includegraphics[width=\linewidth]{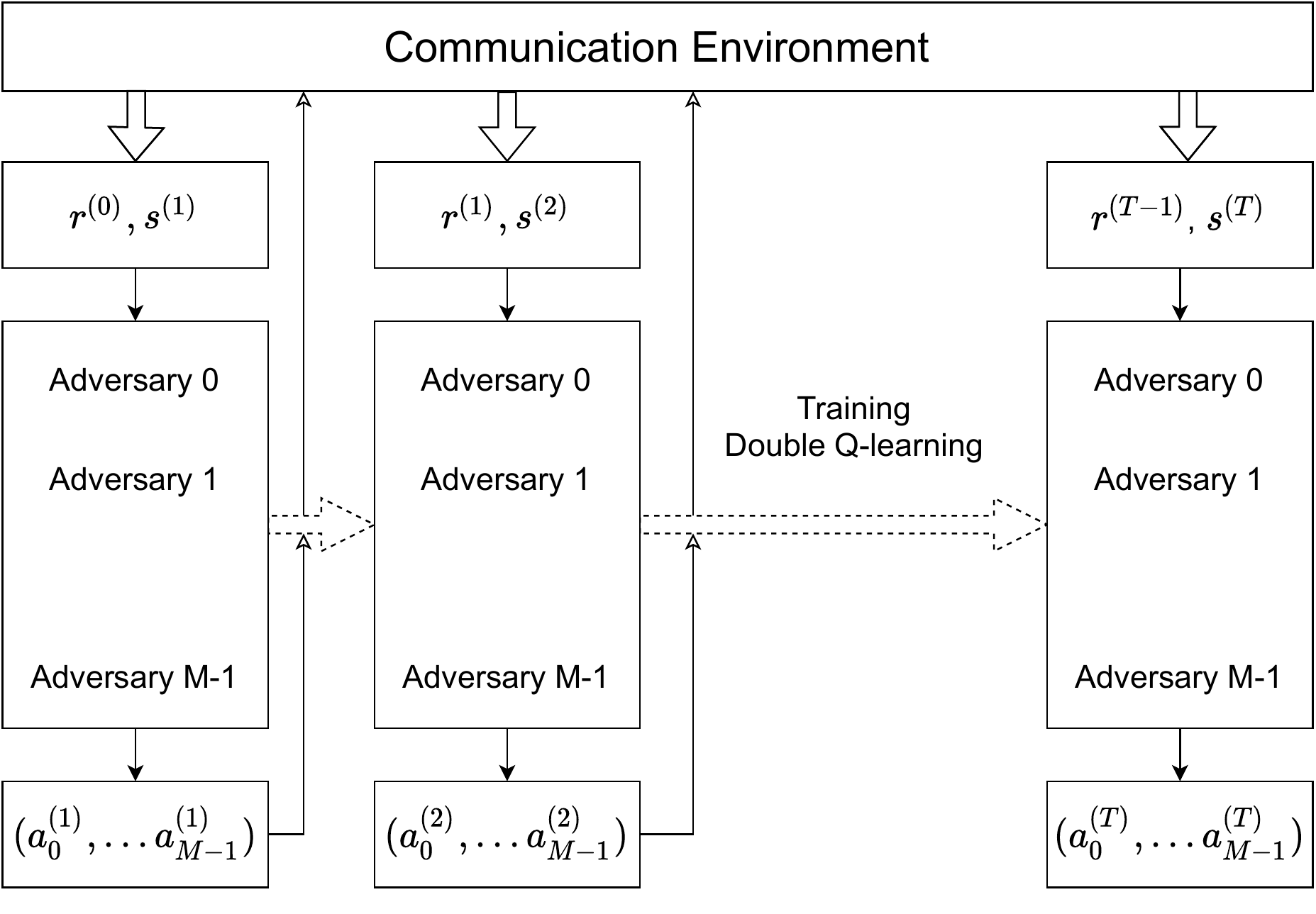}
    \caption{Flow of distributed learning distributed execution MAAS at time $t$. Each adversary $j$ chooses its action independently according to the observed state and all adversaries attack simultaneously. The corresponding SINR values are used for the individual reward of each adversary and as the state for time $t+1$. Training is done separately on each adversary.}
  \label{fig:MAAS}
\end{figure}

\textbf{Learning.} For training of the MAAS, we follow the distributed learning paradigm in which there is no central entity to coordinate the information exchange between adversaries. in addition, all adversaries interact with the environment simultaneously and can observe the SINR values sequentially. They also select their actions independently from each other. Each adversary $j$ is equipped by an experience memory $MEM_j$ for storing the adversary's experience for faster learning, and a pair of actor network $Q_j^{actor}$ and target network $Q_j^{target}$) which are initialized randomly at the beginning of communication. At the very beginning ($t=0$), adversaries make random actions. At any other time $t>0$, each adversary selects its action $a_j^{(t)}$ using its actor network with the current state $s^{(t)}$ (the SINR value from the last time step, $t-1$) as input. The sequence of SINR values is also used for computing individual rewards $r_j^{(t)}$ which is used to train actor networks and target networks $Q_j^{(actor)}$ and $Q_j^{(target)}$ (See figure~\ref{fig:MAAS} and algorithm~\hyperref[alg:MAAS]{1} for details of the learning algorithm and training process). 

\begin{algorithm}[h]
\KwInput{$\gamma,$ $h_s, h_j, \epsilon_j, decay_j, \alpha_j, j \in {0,2,\ldots,M-1}$ $MEM_j(\text{experience storage of adversary j})$ }
\KwOutput{$Q_j^{actor}$}

\tcc{Initialization}
\For{$j=0,2,...,M-1$}{
Initialize $Q_j^{actor}, Q_j^{target}$ \\
Choose $a_j^{(0)}$ as random action
}
$SINR^{(0)}$ is computed based on $[a_0^{(0)},a_2^{(0)},...,a_{M-1}^{(0)}]$ \\
$s^{(1)} = SINR^{(0)}$ \\

\For{$t=1,2,3,\ldots$}{

\tcc{Adversaries choose action}
\For{$j=0,2,...,M-1$}{
\If{$uniform(0,1)>\epsilon_j$}{
$a_j^{(t)} = \operatorname*{argmax}_{a_j}Q_j^{actor}(s^{(t)},a_j)$
}
\Else{
Choose $a_j^{(t)}$ as a random action
}
}
$SINR^{(t)}$ is computed based on $[a_0^{(t)},a_2^{(t)},...,a_{M-1}^{(t)}]$ \\
$s^{(t+1)} = SINR^{(t)}$ \\
\tcc{Receive individual reward and storage the experience}
\For{$j=0,2,...,M-1$}{
Compute $r_j^{(t)}$ \\
Storage the experience tuple $(s^{(t)},a_j^{(t)},r_j^{(t)},s^{(t+1)})$ into $MEM_j$
}
\tcc{Training}
\For{$j=0,2,...,M-1$}{
    Sample experience batch $E$ from $MEM_j$\\
    \tcc{for each experience tuple in batch $E$}
    \For{ $e = (s,a,r,s') \in E$}{
    $a^* = \operatorname*{argmax}_{a_j}Q_j^{actor}(s,a_j)$ \\
    $Target = r + \gamma Q_j^{target}(s',a^*)$ \\
    $Loss = \Big{(}Q_j^{actor}(s,a_j)-Target\Big{)}^2$ \\
    Train $Q_j^{actor}$ through gradient descent on $Loss$ 
    }
    \tcc{Weight Shift for Double DQN}
    $Q_j^{target}=\alpha_j Q_j^{actor} + (1-\alpha_j) Q_j^{target}$
}
\tcc{Decreasing the probability of exploration for more exploitation}
$\epsilon_j = \epsilon_j * decay_j$
}
\label{alg:MAAS}
\caption{Multi-Agent Adversary System}
\end{algorithm}

\subsection{Advantages of MAAS}
Compared to the single-agent system, MAAS is more robust to the failure of any adversary due to the counterattacks of anti-adversaries, communication problems, hardware issues, etc. This is because of the nature of distributed training and executing of agents: if one or more adversaries fail to operate, others can continue their training or execute independently without the failure of the whole system. Furthermore, multi-channel communication in SAA may become a major challenge. The action space of a single adversary grows exponentially with the number of channels it can choose (curse of dimensionality~\cite{COD}). Suppose there are in total $N$ channels and $K$ power levels to choose from, and an adversary is allowed to select $X$ channels at each time step. Therefore, the size of action space is given by $(NK)^X$. If $N=15$, $K=5$ and $X=4$, then the size of action space is $31,640,625$ which is already beyond the capability of most powerful computational resources, not to mention that most adversary devices are small in memory and computation power. Exponentially growing action space implies exponentially more experience to search; thus, exponentially longer time to learn. By distributing the possibility of $X$ channels among multiple agents such that each agent can select less than $X$ channel at each time, MAAS reduces the exponential growth of the action space. For example, if each agent in MAAS is allowed to choose only one channel at each time step, the complexity of the action space will be linear, $O(NK)$, leading to a more efficient and stable learning process. We show this empirically in the experimental section.

\section{Simulations}\label{sec:simulations}
In this section, we conduct extensive simulation studies to show the performance of the proposed MAAS in both single-channel and multiple-channel wireless communication. In each communication scenario, we consider four types of transmitters: (1) Constant sender, (2) Sweep sender, (3) Autoregressive-type (AR-type) sender, and (4) Pulse sender. For each type of sender, MAAS is compared with four benchmark adversaries: (a) Random adversary, (b)Greedy adversary, (c) Single RL adversary that attacks only one channel (SSRL), and (d) Single RL adversary that attacks multiple channels (MSRL). We present more details in the following sections. 
\subsection{Sender Strategies}
Let $N$ denote the total number of available channels, we consider a complementary set of sender strategies in simulations. The first choice is constant sender, which represents the most straightforward strategies. Then, we consider strategies that change with a deterministic pattern that is the sweep sender. Next, the pulse sender simulates the communication environment with abrupt changes. Last, but not least, we use the AR-type sender to represent complicated strategies with extra randomness. Each type of sender selects the channel for signal transmissions by the mechanisms described below.
\begin{enumerate}[(1)]
    \item \textbf{Constant sender} The simplest and easiest strategy of transmission is to stay at one frequency all the time.
    \begin{equation}
    C_{S}^{(t)}=C_{S}^{(0)},
    \end{equation}
    where $C_{S}^{(0)}$ is chosen uniformly from all available channels at initial time $t=0$. 
    \item \textbf{Sweep sender} The choices of channels sweep among all available channels through time. \begin{equation}
        C_{S}^{(t)}=t\%N.
    \end{equation}
    \item \textbf{Pulse sender} We expect this type of sender to evaluate the jammer capability of tracking the abrupt changes.
    \begin{equation}
        C_{S}^{(t)}=\left\{ 
        \begin{array}{ll}5,\; &\text{if}\; t\%N \leq 2\\ 1, \; &o.w.
        \end{array}\right. 
    \end{equation}
    \item \textbf{AR-type sender} On the contrary to the senders above that with deterministic schemes of frequency hopping. The AR-type sender investigates the impact of a non-stationary communication environment.\begin{equation}
        C_{S}^{(t)}=\left\{
        \begin{array}{ll}
        C_{S}^{(t-1)}+i\%N,\;&\text{if} \; C_{S}^{(t-1)}\%2=0\\
        C_{S}^{(t-1)}-i\%N,\;&\text{if} \;C_{S}^{(t-1)}\%2=1\\
        X_t \in \{1, N\},\;&\text{if} \;C_{S}^{(t-1)}>N\\
        Y_t \in \{1, N\},\;&\text{if} \;C_{S}^{(t-1)}<1\\
        \end{array}\right.
    \end{equation}
    where $X_t$ and $Y_t$ are binary random variables with the probabilities $p(X_t=1)=0.1$ and $p(Y_t=1)=0.9$ respectively. At time $t=0$, $C_{S}^{(0)}$ is initialized by an uniformly random choice from all available channels.
\end{enumerate}
For each type of senders, the sender's power $P_S$ is assumed to be fixed throughout the communication.

\subsection{Benchmark Adversaries}
We compare the attacking performance of the proposed MAAS with four benchmark adversaries. Let $M$ denote the number of adversaries in MAAS. The schemes of these adversary attacks are described below.
\begin{enumerate}[(a)]
    \item \textbf{Random} The random adversary interferes by choosing the channel index and allocating power level randomly at each time. 
    \item \textbf{Greedy} The greedy adversary records the history rewards of each action in each iteration. Then, it performs jamming by greedily choosing the action (including channel selection and power allocation) with the highest averaged history rewards. Note that the greedy adversary is a variant of a multi-armed bandit agent.
    \item \textbf{SSRL} The single-channel and single-agent RL (SSRL) adversary is a learning-based intelligent jammer. SSRL performs jamming on one channel at each time.
    \item \textbf{MSRL} The multi-channel and single-agent RL (MSRL) is an intelligent jammer as well. However, in contrast to SSRL, MSRL can produce interference on multiple channels at each time.
\end{enumerate}
We denote the proposed MAAS as the multi-channel and multi-agent RL (\textbf{MMRL}) in all following tables and figures.

\subsection{Simulation Results}

In the simulation studies follows, we set up a synthetic communication environment. There are $N=5$ available channels and four levels of powers $P_S\in \{0, 1, 3, 5\}$. In the single-channel communication system, we fix the power of sender $P_S=5$. While in the multi-channel communication system, the senders can send signals through two channels with the corresponding power $P_S=[1,5]$. The channel gains $h_j$, $h_s$ and cost of power $C_{\text{power}}$ are set to be $1$ without loss of generality.

As for the RL-type adversaries, we set the initial exploration rate $\epsilon = 1$ and the decay rate as $0.996$. During training, the Adam~\cite{Kingma} optimizer is applied with a learning rate of $0.0001$ and no weight decay. All RL-type agents are trained in 2000 epochs. To avoid the uncertainty caused by randomness in the training procedure and random initialization, we report the averaged experimental results over multiple runs for each communication scenario. 

\begin{figure}[h]
    \centering
    \includegraphics[width=\linewidth]{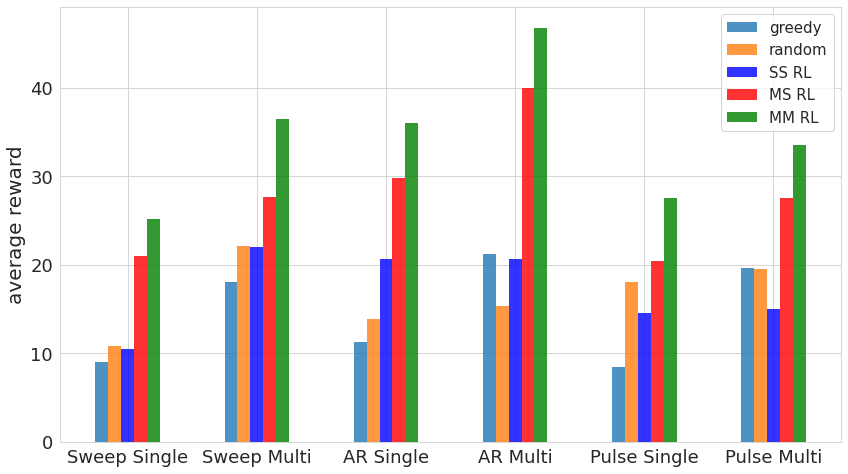}
    \caption{The averaged rewards of various adversaries in different scenarios. The figure shows that MAAS consistently gains over the other adversaries. In particular, it outperforms MSRL in all cases even though we enforce its total power to be the same as the MSRL.}
    \label{figure:rewards}
\end{figure}

Table~\ref{table:successrate} and Figure~\ref{figure:rewards} illustrate the overall performance of adversaries in different scenarios. As shown in Table~\ref{table:successrate}, MAAS has a consistently increased success rate of attacks significantly comparing to all benchmark adversaries regardless of threshold $\tau$.

\begin{table}[h]
  \vspace{-0.5cm}
  \caption{The averaged success rates (\%) of various adversaries on different scenarios. $\tau$ is the threshold for identifying the success of attacks.}
  \label{table:successrate}
  \begin{tabular}{rlllll}
    \toprule
     & \textit{Random} & \textit{Greedy} & \textit{SS RL} &\textit{MS RL} &\textit{MM RL} \\ 
    \midrule
    \multicolumn{6}{c}{\underline{Sweep Single-Channel}} \\
    $\tau=0.2$&46.2 & 39.9 & 31.4 & 72.5 & \textbf{79.1} \\
    $\tau=0.5$&65.5 & 39.9 & 31.4 & 72.5 & \textbf{79.1} \\
    \midrule
    \multicolumn{6}{c}{\underline{Sweep Multi-Channel}} \\
    $\tau=0.2$&49.1 & 40.5 & 33.5 & 49.4 & \textbf{65.9} \\
    $\tau=0.5$&62.3 & 71.2 & 67.7 & 77.2 & \textbf{84.5} \\
    \midrule
    \multicolumn{6}{c}{\underline{Autoregressive Single-Channel}} \\
    $\tau=0.2$&36.7 &57.0 &35.1 &66.6 &\textbf{75.3} \\
    $\tau=0.5$&47.7 &57.0 &35.1 &66.6 &\textbf{75.3} \\
    \midrule
    \multicolumn{6}{c}{\underline{Autoregressive Multi-Channel}} \\
    $\tau=0.2$&30.7 &34.5 &25.1 &49.3 &\textbf{62.3} \\
    $\tau=0.5$&46.5 &67.2 &71.5 &72.5 &\textbf{81.5} \\
    \midrule
    \multicolumn{6}{c}{\underline{Pulse Single-Channel}} \\
    $\tau=0.2$&48.0 &40.0 &53.5 &82.4 &\textbf{93.2} \\
    $\tau=0.5$&65.1 &40.0 &53.5 &82.4 &\textbf{93.2} \\
    \midrule
    \multicolumn{6}{c}{\underline{Pulse Multi-Channel}} \\
    $\tau=0.2$&64.4 &60.1 &66.4 &87.4 &\textbf{94.4} \\
    $\tau=0.5$&81.2 &60.1 &66.4 &87.4 &\textbf{94.4} \\
    \bottomrule
  \end{tabular}
\end{table}
 
\subsubsection{Constant Sender}
\begin{figure}[h]
  \centering
  \begin{subfigure}{0.40\textwidth}
    \includegraphics[width=\linewidth]{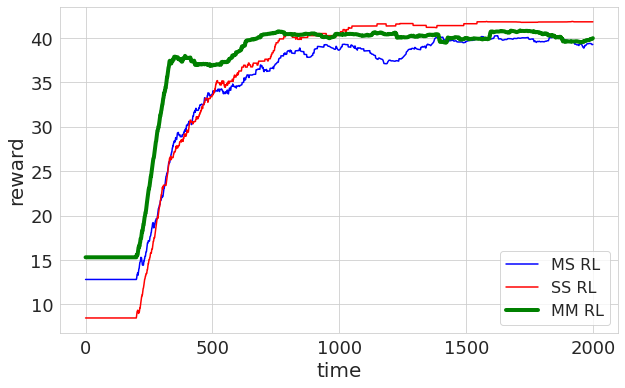}
  \end{subfigure}
  
  \begin{subfigure}{0.40\textwidth}
    \vspace{-0.6cm}
    \hspace*{0.005cm}
    \includegraphics[width=\linewidth]{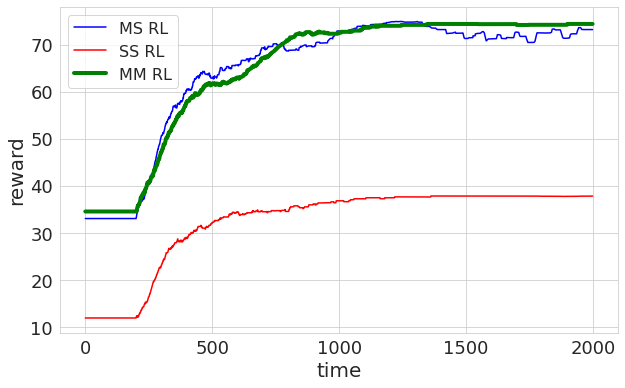}
  \end{subfigure}
   \vspace{-0.3cm}
   \caption{The averaged rewards of RL-type adversaries on constant sender scenarios. On top, the figure shows the performance of attacks in single-channel communications. At the bottom, it shows the performance in multi-channel communications.}
  \label{fig:pp_constant}
\end{figure}

While the constant sender is the simplest sender for adversaries, performance on a constant sender can be an excellent indication of the learning process of different RL adversaries. It is apparent from Figure~\ref{fig:pp_constant} that all three RL-type adversaries converge to the theoretical optimum. In particular, MAAS converges significantly fast than the others. This efficient learning behavior can lead to overall better performance of jamming, as shown in Table~\ref{tab:pp_constant}. On the other hand, Figure~\ref{fig:pp_constant} provides the evidence of a fair comparison in our experiments, that is, the multi-channel RL-type adversaries having a similar optimal reward function as it of the single-channel RL-type adversary. It is beneficial from the reward function we designed, and therefore the optimal reward value for RL agents only has a minuscule increase with more channel capability. 

\begin{table}[h]
    \caption{The averaged success rates (\%) of RL-type adversaries on constant sender scenarios. $\tau$ is the threshold for identifying success of attacks.}
    \label{tab:pp_constant}
    \begin{tabular}{rlll}
    \toprule
    \textit{} & \textit{SS RL} &\textit{Single RL} & \textit{Multi RL}\\
    \midrule
    \multicolumn{4}{c}{\underline{Single-Channel}} \\
     $\tau = 0.2$&  87.5 & 90.7 & \textbf{94.1} \\
     $\tau = 0.5$&  87.5 & 90.7 & \textbf{94.1} \\
    \midrule
    \multicolumn{4}{c}{\underline{Multi-Channel}} \\
     $\tau = 0.2$&  44.9 & 90.1 & \textbf{92.7} \\
     $\tau = 0.5$&  44.9 & 90.1 & \textbf{92.7}\\
    \bottomrule
    \end{tabular}
\end{table}

\subsubsection{Sweep Sender.} 
\begin{figure}[h]
  \centering
  \begin{subfigure}{0.40\textwidth}
    \includegraphics[width=\linewidth]{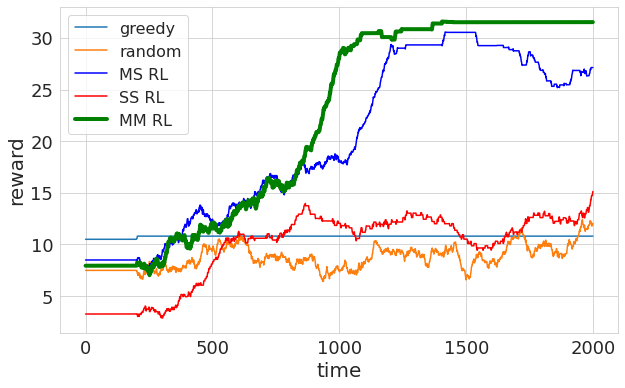}
  \end{subfigure}
  \begin{subfigure}{0.40\textwidth}
    \vspace{-0.6cm}
    \hspace*{0.005cm}
    \includegraphics[width=\linewidth]{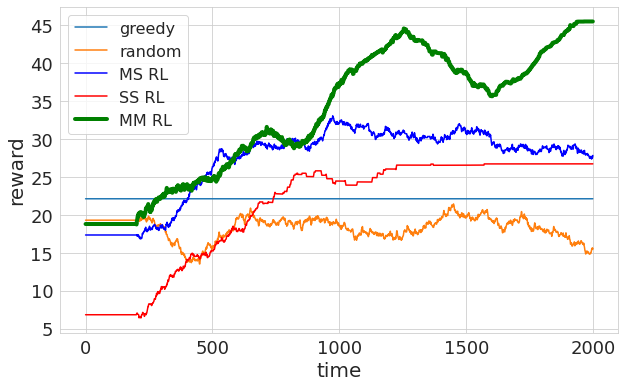}
  \end{subfigure}
  \vspace{-0.3cm}
   \caption{The averaged rewards of various adversaries agents on sweep sender scenarios. On top, the figure shows the performance of attacks in single-channel communications. At the bottom, it shows the performance in multi-channel communications.}
  \label{fig:pp_sweep}
\end{figure}

The most exciting aspect of Figure~\ref{fig:pp_sweep} is the clear demonstration of the training advantage of MAAS, especially in complicated communication scenarios. For example, in the single-channel scenario, the top half of Figure~\ref{fig:pp_sweep} shows that the learning of MAAS is stable, and the convergence of MAAS is fast. At the same time, from the bottom half of Figure~\ref{fig:pp_sweep}, it is observed that MAAS has immense rewards by escaping from the local minimum to which a single agent adversary may converge. 

\subsubsection{AR-type Sender.}
\begin{figure}[h]
  \centering
  \begin{subfigure}{0.40\textwidth}
    \includegraphics[width=\linewidth]{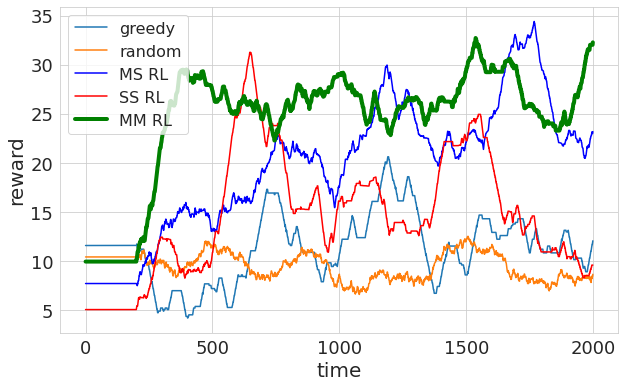}
  \end{subfigure}
  \begin{subfigure}{0.40\textwidth}
    \vspace{-0.6cm}
    \hspace*{0.005cm}
    \includegraphics[width=\linewidth]{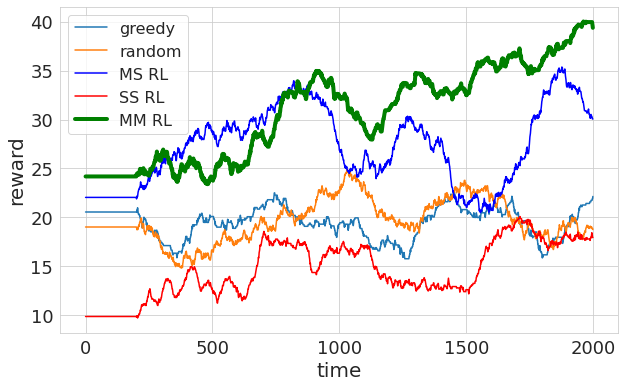}
  \end{subfigure}
  \vspace{-0.3cm}
   \caption{The averaged rewards of various adversaries agents on autoregressive sender scenarios. On top, the figure shows the performance of attacks in single-channel communications. At the bottom, it is the performance in multi-channel communications.}
  \label{fig:pp_ar}
\end{figure}

The AR-type sender is the most complex and challenging sender of all synthetic cases. Nevertheless, MAAS still has a better performance than other types of adversaries facing such a challenge, as shown in Figure~\ref{fig:pp_ar}. For example, MAAS maintains itself in a high-reward region in single-channel communications. Moreover, MAAS tends to converge to a higher reward region in multi-channel communications.

\subsubsection{Pulse Sender.}
\begin{figure}[h]
  \centering
  \begin{subfigure}{0.40\textwidth}
    \includegraphics[width=\linewidth]{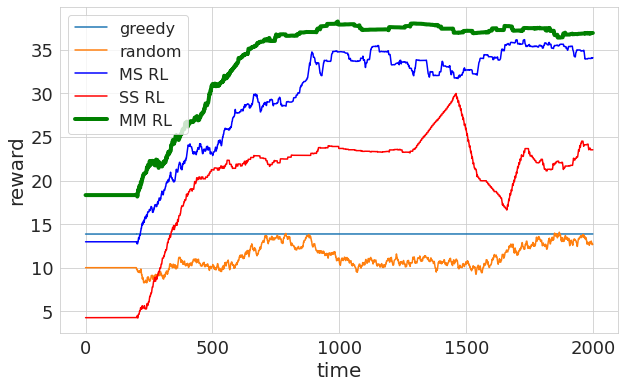}
  \end{subfigure}
  \begin{subfigure}{0.40\textwidth}
    \vspace{-0.6cm}
    \hspace*{0.005cm}
    \includegraphics[width=\linewidth]{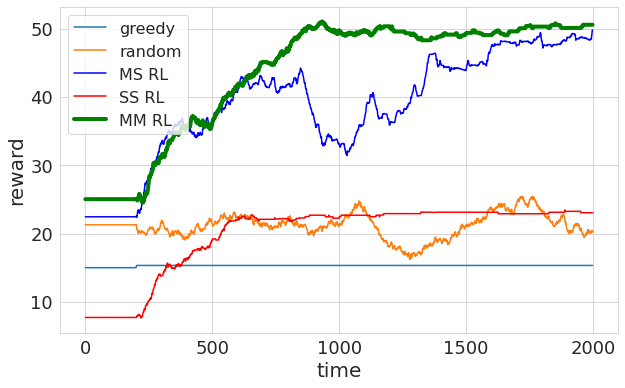}
  \end{subfigure}
  \vspace{-0.3cm}
   \caption{The averaged rewards of various adversaries agents on pulse sender scenarios. On top, the figure shows the performance of attacks in single-channel communications. At the bottom, it is the performance in multi-channel communications.}
  \label{fig:pp_pulse}
\end{figure}

For the pulse sender scenarios, where the communication environment changes abruptly, MAAS has demonstrated its characteristic of efficient learning and stability again. In addition, it has shown faster convergence to optimal and stays stably in high-reward regions in single-channel and multi-channel communications (shown in Figure~\ref{fig:pp_pulse}). Also, Table~\ref{table:successrate} indicates that the success of MAAS is solid. That is, the success rate of MAAS does not change for different values of threshold $\tau$. In fact, random jammers is the only one suffering from this issue. 

\section{One Step Forward: Learning under Real Communication Scenario}
The previous experiments demonstrate the paradigm of the performance of RL-type adversaries in different synthetic communication systems. However, in real-world scenarios, the channel gains $h_s$ and $h_j$ are random variables, while the previous simulation studies assume the constant values of $h_s$ and $h_j$. In this section, we provide some experiments that show that even with random channel gains, the performance of the proposed MAAS remains close to the case of the previous experiments using the constant channel gains. 

We first experiment by setting $h_s$ and $h_j$ as Rayleigh random variables~\cite{RayleighChannel}. Therefore, the corresponding SINR ($SINR_{real}$) is calculated as follows:
\begin{align}
    SINR_{real}^{(t)} &= \frac{P_{S}^{(t)} * h_S * \beta_S }{n+\sum_{j=1}^{N}P_j*h_j*\beta_j*I(C_j^{(t)}=C_S^{(t)})},
\end{align}
where $\beta_S$ and $\beta_j \sim Rayleigh(0, \sqrt{0.5})$. We denote the SINR defined in equation (\ref{sinr}) as $SINR_{cons}^{(t)}$ for the use of constant values of the channel gains. We trained the MAAS with both $SINR_{real}$ and $SINR_{cons}$. we kept all the experimental settings, like the choices of hyper-parameters, the same to ensure that our comparison was fair. Also, we ran our experiments for 8 Monte Carlo trials and reported the average over all the trials. The t-test is then applied to test the difference between the averaged SRAs with degrees of freedom as $7$.  As shown in Table~\ref{table:realSINR}, the results indicate that the MAAS framework is robust to even using random channel gains as MAAS achieves almost the same SRA as the experiment with $SINR_{cons}$.


\begin{table}[h]
  \caption{The averaged SRAs (\%) of MAAS trained with/without raw SINR. (Standard deviations are in the bracket.) P-value closer to 0 indicates more likelihood of significant difference while p-value closer to 1 indicates no significant difference between two averaged SRAs.}
  \label{table:realSINR}
  \begin{tabular}{rllll}
    \toprule
    \textit{} & \textit{Constant} & \textit{Sweep} & \textit{AR-Type} &\textit{Pulse} \\ 
    \midrule
    \multicolumn{5}{c}{\underline{Single-Channel $\tau=0.5$}} \\
    $SINR_{real}$& 93.4(7.4) & 78.2(4.0) & 77.3(6.7) & 93.6(2.4) \\
    $SINR_{cons}$ & 93.1(6.9) & 78.4(4.3) &76.4 (5.9) &94.1(2.6) \\
    p-value & 0.934 &0.925 &0.78 & 0.695 \\
    \midrule
    \multicolumn{5}{c}{\underline{Multi-Channel $\tau=0.5$}} \\
    $SINR_{real}$& 92.5(6.6) & 84.5(4.3) & 81.2(7.2) & 94.5(2.9) \\
    $SINR_{cons}$ & 92.1(6.2) & 85.3(4.8) &81.7 (7.5) &93.9(2.2) \\
    p-value & 0.902 & 0.731 & 0.894 & 0.649 \\
    \bottomrule
  \end{tabular}
 \end{table}

\section{Conclusion}
We have proposed a MARL-based multi-agent adversary system (MAAS) along with a system model for multi-channel communication. We have also defined an appropriate reward function for multi-agent adversaries. MAAS has shown an outstanding performance compared to various baselines, even with power constraints and without partial observability. By distributing actions to multiple agents, MAAS can learn more efficiently and stably than single-agent adversary (SAA) systems. This is due to the linear complexity of the action space in MAAS. The proposed MAAS has its value in active defense as well as understanding the behaviors of multi-agent RL adversaries in designing the defense mechanisms. 

\section{Acknowledgement}
This work was supported in part by Air Force Research Lab Award \#FA 8750-20-2-0504.

\bibliographystyle{ACM-Reference-Format} 
\bibliography{mybib}


\end{document}